# Wehnelt Photoemission in an Ultrafast Electron Microscope: Stability and Usability

Simon A. Willis[1,2], Wyatt A. Curtis[1,2], and David J. Flannigan[1,2,*]

[1]*Department of Chemical Engineering and Materials Science, University of Minnesota, 421 Washington Avenue SE, Minneapolis, MN 55455, USA*

[2]*Minnesota Institute for Ultrafast Science, University of Minnesota, Minneapolis, MN 55455, USA*

**Abstract:** We tested and compared the stability and usability of three different cathode materials and configurations in a thermionic-based ultrafast electron microscope: (1) on-axis thermionic and photoemission from a custom 100-μm diameter $LaB_6$ source with graphite guard ring, (2) off-axis photoemission from the Ni aperture surface of the Wehnelt electrode, and (3) on-axis thermionic and photoemission from a custom 200-μm diameter polycrystalline Ta source. For each cathode type and configuration – including the Ni Wehnelt aperture, we illustrate how the photoelectron beam-current stability is deleteriously impacted by simultaneous cooling of the source following thermionic heating. Further, we demonstrate usability via collection of parallel- and convergent-beam electron diffraction patterns and by formation of optimum probe size. We find that usability of the off-axis Ni Wehnelt-aperture photoemission is at least comparable to on-axis $LaB_6$ thermionic emission, as well as to on-axis photoemission (the heretofore conventional approach to UEM in thermionic-based instruments). However, the stability and achievable beam currents for off-axis photoemission from the Wehnelt aperture were superior to that of the other cathode types and configurations, regardless of the electron-emission mechanism. Beam-current stability for this configuration was found to be ±1 % (one standard deviation from the mean) for 70 minutes (longest duration tested), and steady-state beam current was reached within the sampling-time resolution used here (~1 s) for 15 pA beam currents (*i.e.*, 460 electrons per packet for a 200 kHz repetition rate). Repeatability and robustness of the steady-state condition was also found to be within ±1 % of the mean. We discuss the implications of these findings for UEM imaging and diffraction experiments, for pulsed-beam damage measurements, and for practical switching between optimum conventional TEM and UEM operation within the same instrument.

*Author to whom correspondence should be addressed.
Email: flan0076@umn.edu
Office: +1 (612) 625-3867

**Introduction**

Femtosecond (fs) laser-based ultrafast electron microscopy (UEM) employs photoemission from a source in the electron gun region [1–4]. All three of the main transmission electron microscope (TEM) gun types – thermionic, Schottky field emission, and cold field emission – have been shown to produce viable UEM operation with varying degrees of performance that roughly trend with typical conventional operation [2,3,5–9]. The first laser-based UEM dedicated to fs pump-probe operation (UEM-1) used a standard Wehnelt-based LaB$_6$ thermionic electron gun (TEG) [10]. The second-generation laser-based fs-centric instrument (UEM-2) used a field-emission gun (FEG) equipped with a LaB$_6$ source [11]. The development and use of Schottky and cold FEGs for fs laser-based UEM operation is more recent, driven largely by a desire for higher brightness and improved coherence at the expense of achievable electron-packet size (*i.e.*, beam current) and, thus, repetition-rate ($f_{rep}$) flexibility relative to LaB$_6$ TEGs.

In fs laser-based UEM (as with all ultrafast electron-based measurement techniques), preservation of high temporal resolution comes at the expense of electrons per packet due to electron-electron repulsion [12–14]. The resulting low beam currents can be offset by increasing the $f_{rep}$ [15]. This is because laser-based-UEM average photoelectron (*pe*) beam current, $\overline{I_{pe}}$, is directly proportional to $f_{rep}$: $\overline{I_{pe}} = \left[\left(\frac{E_{lp}}{hv} \cdot \eta(hv, T)\right) \cdot CE\right] \cdot e \cdot f_{rep}$ [16]. Here, $E_{lp}$ is the energy per laser pulse incident on the photoemitter, $hv$ is the incident photon energy, $\eta$ is the photon-energy- and temperature-dependent photoelectron-source quantum efficiency defined as the ratio of the number of photoelectrons per packet to the number of incident photons of energy $hv$ per laser pulse [17], $CE$ is the collection efficiency defined as the ratio of the number of emitted photoelectrons passing through the X-ray aperture and entering the illumination system to the total number emitted at the source, and $e$ is the fundamental charge. (Note the specific value of $CE$ depends upon how it is defined – one can choose other points along the optic axis to define *CE*, such as the specimen position [18].)

Operating at elevated $f_{rep}$, however, likely limits the range of phenomena that can be probed [19]. Low $f_{rep}$, high-resolution UEM (HR-UEM) is possible but requires long acquisition times and presumably high instrument and lab stabilities [20]. Further, UEM experiments often consist of multiple individual acquisitions that, when taken altogether, can span several hours or more for a single time-scan. Under such conditions, photoelectron beam current would ideally be stable spanning such timescales so that re-heating or flashing the source during experiments is avoided and so that one need not perform potentially non-representative corrections to data obtained with decaying beam current. Stable photoemission is also desirable for conducting pulsed-beam radiation-damage experiments with laser-based UEM [21,22].

Despite the clear benefits, there is presently a dearth of data on photoelectron beam-current stability in UEMs. However, anecdotal evidence and table-top measurements of common source materials suggest long-term stability spanning hours is generally poor, with beam-current decay being roughly similar to that of thermionic and field emission in conventional TEMs [6,23]. Indeed, such decays were a motivating factor for using second-harmonic light from a Ti:sapphire fs oscillator and a LaB$_6$ source for the initial configuration of UEM-1 ($hv$ = 3.2 eV compared to $\Phi_{LaB_6} \cong 2.7$ eV) [10]. It was hypothesized that matching $hv$ and work function ($\Phi$) would improve stability and coherence, though again at the expense of electrons per packet owing to the reduced quantum efficiency [24,25]. (With UEM-1, the resulting low number of electrons per packet was compensated for by using a laser $f_{rep}$ of up to 80 MHz [10,19,26,27].) However, while UEM-1 beam-current stability was not documented, measurements with table-top devices indicate decays are still present when using $hv$ = 3.16 eV [23].

Here we characterize the stability and usability of fs photoelectron beams generated from three different source materials in a 200 kV TEG-based UEM equipped with a conventional Wehnelt triode. Of particular note, we demonstrate the usability and we quantify the stability of photoelectron beams from the Ni Wehnelt-aperture anode-facing surface. Using fs laser pulses with $h\nu$ = 4.8 eV, *we find that stability and the achievable steady-state beam current from the Ni aperture are significantly improved compared to LaB$_6$ photoemission in the same instrument*. We show that diffraction-pattern quality and achievable probe size for Wehnelt aperture-surface photoemission is at least comparable to both photo and thermionic emission from a custom truncated LaB$_6$ source with diameter approximately equal to that of the probe-laser spot size ($e^{-2}$ Gaussian width of 80 µm). With respect to versatility and usability of TEG-based UEMs, we find that the off-axis Wehnelt-aperture photoemission allows one to conduct conventional TEM with the same instrument using an on-axis LaB$_6$ cathode optimized for high-quality thermionic emission (*e.g.*, 16-µm flat diameter set 0.35 mm back from the aperture plane [28]). Because stability was improved for photoemission from the metallic aperture surface ($\Phi_{Ni} \cong$ 5.0 eV [29]), we also quantified the photoemission properties and behavior of a custom 200-µm diameter polycrystalline (*pc*) tantalum cathode [$\Phi_{pc-Ta} \cong$ 4.25 eV]. Again, both stability and steady-state beam current relative to LaB$_6$ photoemission were improved, and the photobeam properties were again at least as good. However, the on-axis *pc*-Ta source did not outperform off-axis photoemission from the Ni Wehnelt aperture. It also did not provide a stable conventional thermionic beam, thus negating any practical benefits with respect to switching between TEM and UEM operation.

**Methods**

All experiments were performed on a Thermo Fisher/FEI Tecnai Femto 200 kV UEM/TEM located in the Ultrafast Electron Microscopy Lab at the University of Minnesota [3]. Typical base pressure in the electron-gun region during all measurements was on the order of $10^{-7}$ Torr. The detector used to measure beam current was a Gatan OneView 16 MP CMOS camera. The manufacturer calibration was checked with a Faraday cup and picoammeter [21]. The base microscope is a Tecnai $G^2$ T20 200 kV TEM equipped with a standard Wehnelt triode. Modifications for UEM operation consist of two optical periscopes incorporated into the side of the TEM column. The microscope is interfaced with a Light Conversion PHAROS 6W diode-pumped solid-state fs pulsed laser (Yb:KGW). Pulse duration of the fundamental laser output ($h\nu$ = 1.2 eV) was measured with a Light Conversion GECO scanning autocorrelator to be 240 fs fwhm. The second harmonic ($h\nu$ = 2.4 eV) is generated with a Light Conversion HIRO harmonics module, and the fourth harmonic ($h\nu$ = 4.8 eV) is generated using a BBO after the module (see Ref. [30] for a schematic of the laser table layout). Fourth-harmonic light was used to generate pulsed photoelectron beams for all source types and configurations. All photoemission was driven by single-photon effects, as confirmed by a linear response for beam current vs. laser power. The probe laser spot size on the electron source was estimated to be 80 μm ($e^{-2}$ Gaussian width) by measuring the beam profile (Newport, NP LBP2-VIS2) external to the microscope column and then extrapolating to the source using the final lens focal length, the collimated beam width, and the distance to the source [31]. Probe laser power was also measured external to the column with a Newport NP 919P-003-10 power meter. Measured average power was 21.7 mW for all photobeam experiments (0.1 μJ/pulse), with a resulting incident fluence of 2.2 mJ/cm on the source (calculated using the area from the $e^{-2}$ width and a laser $f_{rep}$ = 200 kHz). Beam current and average

Page 5 of 25

electrons per packet were measured using the calibrated OneView sensor with the beam entirely converged onto the chip.

Three photoelectron gun configurations and geometries were tested and compared and are reported here: (1) on-axis photoemission from a custom 100-μm diameter truncated LaB$_6$ cathode with graphite guard ring (Fig. 1a), (2) off-axis photoemission from a 0.5- and a 1-mm diameter Ni Wehnelt aperture surface (Fig. 1b), and (3) on-axis photoemission from a custom 200-μm diameter truncated *pc*-Ta cathode. For comparison purposes, thermionic emission stability and usability (*i.e.*, probe size and diffraction-pattern quality) were also characterized for the 100-μm LaB$_6$ with graphite guard ring. Thermionic emission from the *pc*-Ta source was also tested, but stability was found to be poor and so is not reported owing to one of our goals being to identify configurations amenable to rapid, convenient, and robust switching between TEM and UEM modes. (Note that we chose *pc*-Ta instead of Ni for the on-axis metal-cathode photoemission and thermionic comparisons owing to the significantly higher melting temperature of Ta.) All cathodes were custom fabricated and supplied by Applied Physics Technologies.

Thermionic emission was carried out above the thermal emission threshold at a typical heat-to value of 30 (per the Tecnai user interface), while photoemission was carried out below the thermionic threshold at heat-to values of 24 (*pc*-Ta), 20 (LaB$_6$), or 0 (LaB$_6$ and the Ni Wehnelt aperture). For reference, onset of observable thermionic emission is found to occur at a heat-to value of approximately 27 in our instrument, indicative of a LaB$_6$ temperature of over 1400 K [17]. The on-axis cathodes were set-back 0.35 mm from the Wehnelt-aperture mid-plane [18,28]. Wehnelt aperture composition was determined to be Ni by conducting energy dispersive X-ray analysis on a cross-sectioned specimen. The analysis also confirmed that the aperture surface was free of contamination, indicating Ni was indeed the photoemitting material. This was confirmed by observing no change in performance after removing, polishing, cleaning, and re-installing the aperture. The Wehnelt apertures were supplied by Thermo Fisher Scientific.

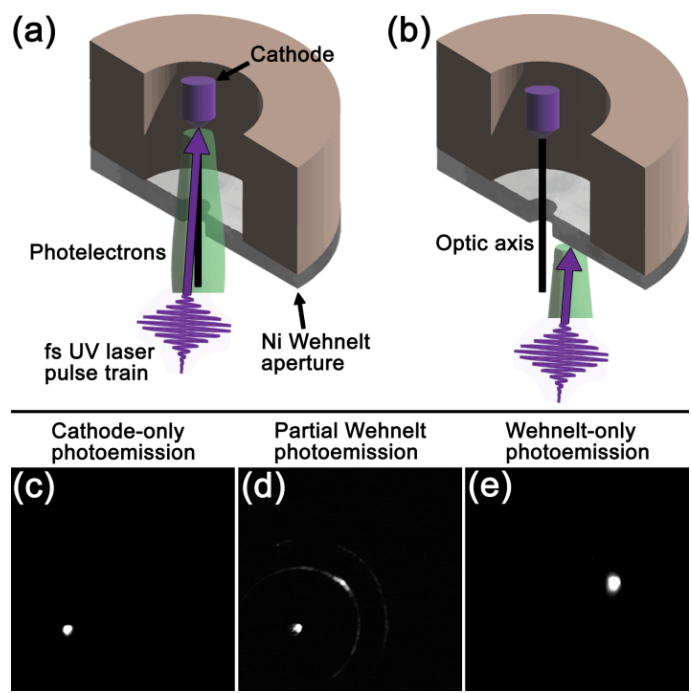

**Figure 1.** Tecnai Femto UEM photoemission geometries studied here. Illustrations of (a) on-axis photoemission from the cathode and (b) off-axis photoemission from the Ni Wehnelt-aperture anode-facing surface. The black vertical lines indicate the system optic axis. The green truncated cones represent general trajectories of the photoelectron packets. The approximate incident direction of the fs UV laser pulse train is also shown. (c) Pulsed-beam image of on-axis photoemission from a custom 100-μm diameter truncated $LaB_6$ cathode with graphite guard ring. (d) Pulsed-beam image of partial Wehnelt photoemission resulting from translation of the incident UV laser away from the on-axis $LaB_6$ cathode position in panel (c). The observed photoemission pattern is due to the UV laser partially striking the inner aperture surface. (e) Pulsed-beam image of off-axis photoemission entirely from the anode-facing Ni Wehnelt-aperture surface.

The laser and electron-beam alignment procedures were as follows. First, the cathode was heated to the point of weak thermionic emission, and electron-beam alignments were performed. Second, the fs laser-pulse train was aligned to the cathode using a mirror housed in a piezo mount in the Tecnai Femto gun periscope. Alignment was optimized by maximizing the beam current via iterative laser-beam translation on the cathode. Third and finally, the cathode heat-to value was reduced to either 24, 20, or 0, depending upon the specific experiment. This produced a purely photoelectron beam that then underwent final optimization alignments using gun shift and gun tilt. A representative image of photoemission from the custom 100-μm $LaB_6$ source is shown in Figure 1c. To generate off-axis Wehnelt-aperture photoemission, the laser-spot position was laterally translated from the $LaB_6$ cathode to the anode-facing Ni aperture surface by adjusting the position of the mirror in the piezo mount. During translation, photoemission from the inner aperture surface was observed, as was still-present but diminished photoemission from the cathode (Fig. 1d). Translation was deemed complete once photoemission was observed emanating entirely from the anode-facing aperture surface (Fig. 1e). Photoelectron beam alignment was then re-optimized.

In UEMs that employ $LaB_6$ cathodes and a TEG, the cathode temperature can be increased but held below the thermal emission threshold in order to increase the photoelectron beam current [17,23]. To test the impact heating has on photoelectron beam stability in our instrument, and to make general comparisons to prior work [23], the following procedure was used. First, the $LaB_6$ cathode was held at a heat-to value of 30 (*i.e.*, above the thermal emission threshold) for at least 15 minutes prior to reduction to sub-thermal-emission values of either 20 or 0 (a heat-to value of 24 was used in the *pc*-Ta experiments). Second, once the reduced heat-to set-point was reached, the laser was un-blocked, and photoelectron beam-current measurements were started (dubbed time zero, $t = 0$). In general, the temperature of the photoemitter is expected to influence beam current due to Fermi-Dirac statistics and due to adsorption of contaminants on the emitting surface [17,32–34]. Such effects manifest in the associated LaB$_6$ stability data reported here as a decaying beam current, wherein the cathode was cooling to ambient conditions while the photoelectron beam current was being measured (see Figs. 2 and 3). Such cooling lowers the probability of occupying states above the Fermi energy and also increases the sticking probability of contaminants, which in turn will increase the effective $\Phi$ of the photoemitter [17,33,34]. (Note that direct, quantitative comparisons to prior work are difficult owing to the complexities of equilibration – see Fig. 3 and Ref. [23], for example.)

Importantly, similar decays in photoelectron beam current for Ni Wehnelt-aperture photoemission also occur if measurements are started before the gun region has cooled completely (see Fig. 4c). As such, in order to ensure that the inherent stability and quality of the Wehnelt-aperture photoelectron beam was tested, the cathode heat-to value was set to 0, and the gun region was allowed to cool and equilibrate to ambient conditions prior to conducting photoelectron-beam measurements. This was done to ensure radiant heating of the Wehnelt aperture by the hot LaB$_6$ cathode had fully dissipated prior to starting the experiment. As will be discussed, *this results in immediate, robust, and prolonged photoelectron-beam stability from the Ni Wehnelt aperture*.

**Results and Discussion**

Shown in Figure 2 is behavior typical of on-axis photoemission from a LaB$_6$ cathode immediately after reducing the heat-to value from above to below the thermal emission threshold. Here, time zero ($t = 0$) denotes when the reduced heat-to value set-point is reached and photoemission is started. Decay in photoelectron beam current and electrons per packet is typical, regardless of the simultaneously-applied sub-thermal-emission heat-to value [23,35]. Also, in addition to enhancing photoemission currents above a certain LaB$_6$ cathode temperature [17], using non-zero heat-to values (*e.g.*, 20) provides a more stable beam, as indicated by the larger time constants from the bi-exponential decay fits [23,35].  Note that the use of a bi-exponential decay fitting function is not meant to suggest a specific mechanism but instead is for comparing temporal behaviors across the different photoemission configurations and settings.  That said, such behavior may indeed be due to convolution of responses (*e.g.*, shifting population distributions and increasing effective $\Phi$ with increasing surface contamination while cooling [17,33,34]).  Beam current and electrons per packet continued to gradually decay for several hours after reducing the heat-to value (here, from 30 to 20 or 0).  The beam-current decay rates are expected to depend on the pressure in the electron-gun region (here, ~$10^{-7}$ Torr) [32–34].  The beam current could not be improved by re-optimizing laser alignment on the LaB$_6$ or by re-optimizing the electron-beam alignment [6], indicating the alignments were stable and robust during the measurements.

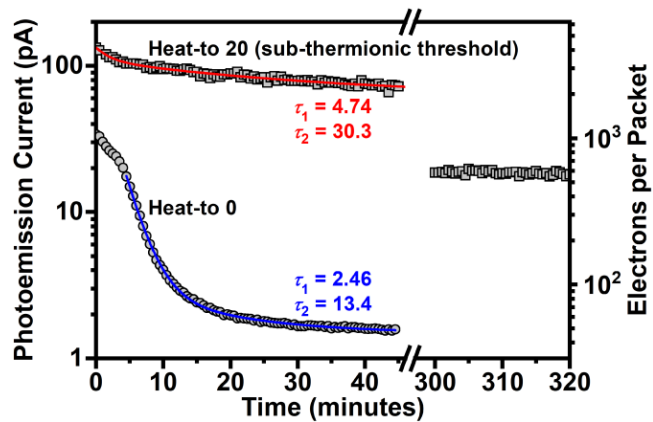

**Figure 2.**  Characteristic temporal (*t*) beam-current behavior for on-axis LaB$_6$ photoemission immediately following thermionic operation.  Behaviors for non-thermionic heat-to values of 20 (squares) and 0 (circles) are shown.  The heat-to value was reduced from 30 to 20 or 0, and measurement of photoemission current commenced immediately thereafter, corresponding to *t* =

Page **10** of **25**

0 minutes. Heat-to 20 and 0 data points up to 45 minutes are the average of three and two separate measurements, respectively (deviation from the mean is equal to or smaller than the data marker size, representing a less than 3% deviation). An example of longer-term behavior is shown for heat-to 20 beginning at 300 minutes. Data up to 45 minutes are fit with a bi-exponential decay function to quantitatively compare behaviors. The fits are labeled with the respective time constants, $\tau_1$ and $\tau_2$, in minutes. (Note that the heat-to 0 data were fit beginning at $t = 4.5$ minutes owing to the anomalous deviation in decay rate between 0 and ~5 minutes.)

The initially relatively high photoelectron beam currents observed immediately following cooling to sub-thermionic-emission thresholds can be recovered by again re-heating the LaB$_6$ cathode to above the threshold (Fig. 3) [17]. In addition to shifting the electron population back to a higher-temperature distribution, the effective $\Phi$ will decrease due to removal of surface contaminants [33,34]. For the experiment summarized in Figure 3, beam current was monitored while cycling between thermionic emission at a heat-to value of 30 and photoemission at a heat-to value of 20. At heat-to 30, thermionic beam current initially rises before roughly plateauing within ~30 minutes of setting the heat-to value [23]. (Note that the mostly repeatable plateauing that occurs approximately 15 minutes before a subsequent jump and second plateauing in thermionic emission current is similar in appearance to the evolution of LaB$_6$ surface oxygen with temperature [34].) The heat-to value was then reduced to 20, reaching the set-point within ~30 seconds. This was immediately followed by unshuttering of the fs UV laser-pulse train and monitoring the resulting photoelectron beam current. As can be seen, the photoelectron beam-current decay follows a similar bi-exponential decay as shown in Figure 2. Note that because a heat-to value of 20 is below the threshold for observable thermionic emission, the measured absolute photoemission beam current at the moment of laser unshuttering is well below the purely thermionic beam current generated with a heat-to value of 30 [23]. Also, the absolute initial photoelectron beam currents are higher here than in Figure 2 (250 pA vs. 130 pA) owing to less time elapsed between heat-to value reduction and initiation of photoemission.

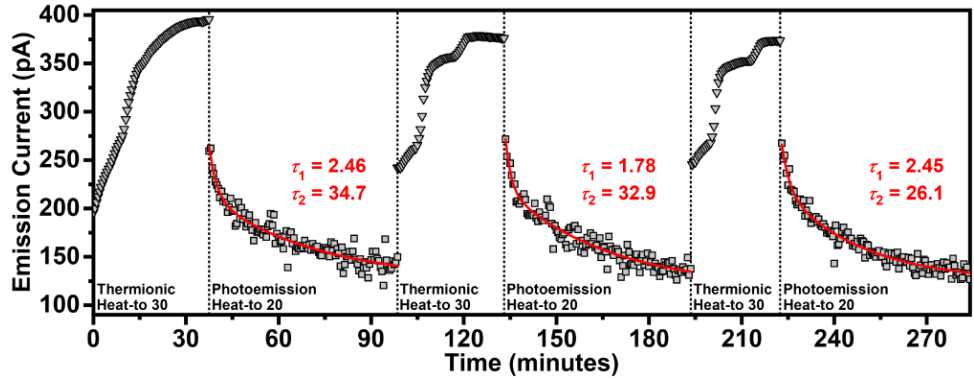

**Figure 3.** Beam-current behavior when cycling between thermionic (heat-to 30; inverted triangles) and photoelectric (heat-to 20; squares) emission from an on-axis 100-µm diameter $LaB_6$ source with graphite guard ring. Photoemission temporal behavior is quantified by fitting the data with a bi-exponential decay function (red). Time constants in minutes are shown above the corresponding data.

Before discussing the results of photoemission from the Ni Wehnelt aperture surface, we briefly address the apparent quantitative repeatability of the $LaB_6$ photoemission beam-current decay shown in Figure 3. Because the time constants of the bi-exponential decay fits are comparable across the three cycles shown, the behavior may be amenable to statistical characterization for the purposes of generating a correction function to be applied to UEM time-series data. This is a practical issue. Instead of attempting to immediately conduct experiments at the onset of photoemission, one could allow the system to reach a quasi-stable condition a few hours after initiation [23]. However, we have found that the duration of the heat-to 30 setting applied prior to initiating photoemission impacts the time constants but not the overall bi-exponential behavior, further confirming the complexity of equilibration. Indeed, this variability led us to pursue a practical, more robust solution to the beam-current decay problem that improves practicality with respect to conducting UEM experiments and switching between pulsed and thermionic modes. Nevertheless, though not the emphasis here, we do not rule out the possibility of successfully generating and applying such a correction function to data obtained under well-controlled and rigorously characterized conditions.

Because the photoemission beam-current decay shown in Figures 2 and 3 is non-ideal with respect to both ultrafast measurements and to pulsed-beam damage studies, we sought alternative geometries and source materials. The goal was to find configurations that provide increased and immediate UEM-operating stability while also preserving optimum thermionic performance and providing rapid and routine switching between TEM and UEM operation. *Accordingly, we found that one can generate viable photoelectron beams from the Ni Wehnelt aperture surface of the TEG in the Tecnai Femto UEM.* This can be done by laterally translating the fs UV laser-pulse train from the on-axis cathode to an off-axis position on the aperture surface (see Fig. 1c-e), followed by re-optimization of the electron-beam alignments owing to the off-axis geometry. Note that photoemission from the extractor surface in a Schottky-FEG UEM instrument has been demonstrated and characterized in terms of energy spread, temporal duration, and brightness [36]. Unfortunately, direct comparisons of performance metrics of different UEM instruments and labs is difficult owing to a combination of a very large number of variables and to the current dearth of statistical data sets generated from large numbers of measurements [16,20]. Thus, the present value of studies such as this one is in assessing viability of specific configurations for accessing certain experimental parameter space. Nevertheless, work such as this also contributes to the growing literature on UEM performance and capability.

Stability and robustness of an electron beam photoemitted from the Ni Wehnelt-aperture anode-facing surface is shown in Figure 4. Note that these data sets are representative of the stabilities and robustness typically observed here for Wehnelt photoemission. For example, the standard deviation in beam current (and electrons per packet) over the span of nearly 70 minutes is on the order of ±1% for an average current of 14.7 pA (*i.e.*, 460 e$^-$/packet) (Fig. 4a). For $hv$ = 4.8 eV, the quantum efficiency ($\eta$) of Ni is ~$10^{-6}$ [29]. Here, $10^{11}$ photons/pulse were incident on the aperture surface, giving $10^5$ photoelectrons per pulse under optimum conditions. This is in reasonable agreement with the measured value here once typical losses of ~100× from source to detector are accounted for [18]. (Note that Ni specimens can display a bivalued $\Phi$, depending upon microstructure, with onset of the lower value occurring at 4.5 eV [29].) For this particular data set, a least-squares linear fit returns a slightly positive slope of 0.5 fA/min. [0.02 (e$^-$/packet)·min.$^{-1}$]. This is likely due to the scatter in the data and not to an actual upward trend in beam current with time. As described in the Methods section, the cathode is not heated during these measurements, and the gun assembly has been allowed to completely cool to ambient conditions prior to acquiring photoelectron beam-current data. Note that $t = 0$ is defined as the moment the fs UV laser-pulse train is unblocked and photoemission begins.

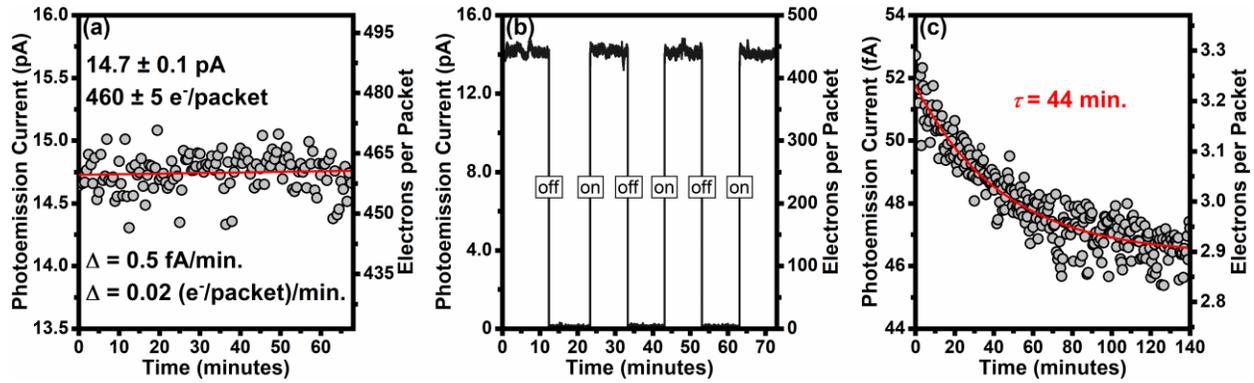

**Figure 4.** Photoemission behavior from the Ni Wehnelt-aperture anode-facing surface. (a) Stability behavior from a thermally equilibrated, ambient-temperature Ni Wehnelt-aperture surface. The average photocurrent and electrons per packet were 14.7 ± 0.1 pA and 460 ± 5 e$^-$·packet$^{-1}$, respectively. Errors are one standard deviation from the average. The red line is a linear least-squares fit of the data. The slope of this line (Δ) is 0.5 fA·min$^{-1}$ or 0.02 (e$^-$·packet$^{-1}$)·min$^{-1}$. The $f_\mathrm{rep}$ was 200 kHz. (b) Demonstration of the immediate response and repeatability of photoemission from the thermally equilibrated, ambient-temperature Ni Wehnelt aperture. The "off" and "on" labels correspond to when the fs UV laser-pulse train was shuttered and unshuttered, respectively. (c) Example of the photoemission behavior from a Ni Wehnelt aperture that is cooling to ambient temperature following heating of the cathode. The data are fit with a single exponential decay function (red). The beam current and electrons per packet are significantly lower compared to the data in (a) owing to sampling at a later time in the total decay. Nevertheless, the observed behavior is typical for a still-cooling electron-gun region.

The data in Figure 4a also shows that the steady-state beam current is reached as soon as the UV laser is trained on the Wehnelt aperture ($t = 0$). To test the robustness and repeatability of this response, a mechanical shutter placed between the laser source and the electron source was alternately opened and closed with a roughly 50% duty cycle and a roughly uniform pulse width of 10 minutes. Figure 4b shows the resulting response of the Ni Wehnelt-aperture photoemission current. With each on-cycle, the photoemission current was observed to be at its steady-state value by the first data-point acquisition (*i.e.*, within 1 second). Further, the overall stability of each on-cycle was the same as that shown in Figure 4a, and the deviation from period to period was within 1% for the series shown. We emphasize here that the data in Figure 4a,b were obtained from an electron gun that had fully equilibrated to ambient conditions. Indeed, the importance of allowing the gun to fully equilibrate to ambient conditions is illustrated in Figure 4c. Here, photoemission from the Ni Wehnelt aperture was tracked soon after the heat-to value for the $LaB_6$ cathode had been reduced from a condition of thermionic emission to zero and while the gun region was still cooling to ambient condition. A non-linear drop in current occurs, qualitatively similar to what occurs for photoemission from $LaB_6$ prior to complete cooling and equilibration [23].

While the stability and robustness of photoemission from the Wehnelt aperture is an improvement relative to photoemission from the on-axis $LaB_6$ source, the off-axis geometry requires determination of usability of the Wehnelt photobeam. Accordingly, we measured the smallest probe size that could be generated, and we acquired diffraction patterns requiring a moderate level of beam coherence (Fig. 5). As shown in Figure 5a, a probe size of approximately 20 nm (fwhm) could be generated from the Wehnelt-aperture photobeam using the Nanoprobe mode of the Tecnai Femto. We deem this result reasonable owing to the 80 μm $e^{-2}$ Gaussian width of the fs UV laser spot size on the aperture. Further, this probe size was the same as that generated by on-axis photoemission from the custom 100-μm diameter $LaB_6$ cathode (Fig. 5b) and by conventional thermionic emission from the same $LaB_6$ cathode (Fig. 5c). Note that the gun alignments were separately optimized for each emission configuration in order to achieve the smallest-possible probe size for each.

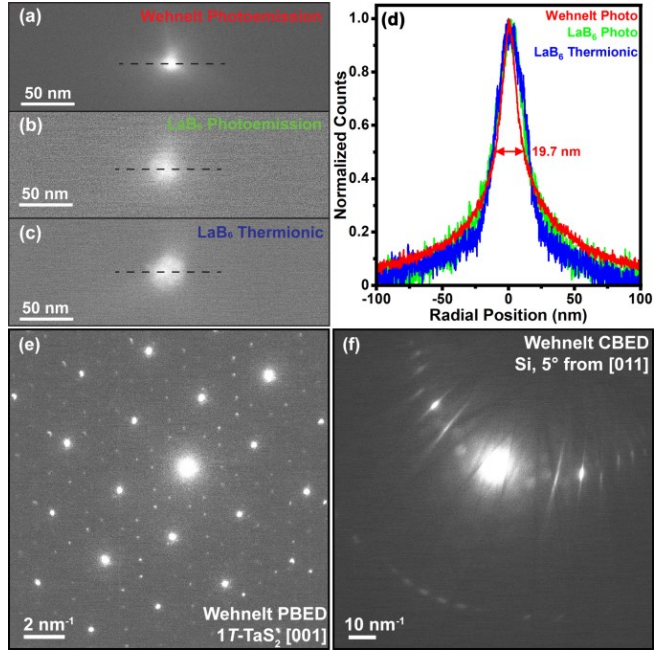

**Figure 5.** Probe size and diffraction-pattern quality of Ni Wehnelt-aperture photoemitted beams. (a-c) Spot-size images of probes formed using the Nanoprobe mode of the Tecnai Femto: (a) a photoelectron beam from the Ni Wehnelt aperture, (b) a photoelectron beam from the on-axis, 100-μm diameter $LaB_6$ source, and (c) a thermionic beam from the same $LaB_6$ source. The UV laser spot size on the aperture and the $LaB_6$ source was 80 μm ($e^{-2}$ diameter). Black dashed horizontal lines mark positions from which the line profiles in (d) were generated. The fwhm was approximately 20 nm for each probe. (e) Parallel-beam electron-diffraction (PBED) pattern from multilayer $1T$-$TaS_2$ obtained along the [001] zone axis using the Wehnelt photoelectron beam. (f) Convergent-beam electron-diffraction (CBED) pattern from Si tilted 5° off the [011] zone axis generated using the Wehnelt photoelectron beam.

Diffraction patterns generated with the Wehnelt-aperture photobeam are shown in Figure 5e,f. Figure 5e is a parallel-beam electron-diffraction (PBED) pattern of multilayer $1T$-$TaS_2$ along the [001] zone axis with the first- and second-order charge-density-wave superlattice spots apparent and resolved. This is noteworthy because the maximum counts of the Bragg spots and the superlattice spots differ by ~100×. Figure 5f is a convergent-beam electron-diffraction (CBED) pattern of single-crystal Si approximately 5° off the [011] zone axis. One can see that the second-order Laue-zone ring and the Kikuchi bands are observable. Taken altogether, the results in Figure 5 show that Wehnelt photoemission is at least comparable in quality and usability to on-axis photoemission from the large $LaB_6$ cathode, despite the off-axis configuration and necessary adjustment of gun shift and gun tilt alignments. That is, it appears that off-axis photoemission from the Wehnelt-aperture surface is an at least viable configuration for UEM operation.

Considering that access to thermionic and photoemission beams and convenient switching between TEM and UEM operation are appealing aspects of Wehnelt-aperture photoemission, we tested the performance of an on-axis, custom 200-μm diameter *pc*-Ta cathode (Fig. 6). Note that Ta sources have been shown to be viable cathodes for fs laser-based UEMs based on TEG TEMs with a Wehnelt electrode [8], and metal cathodes in general are widely used in dedicated ultrafast electron diffraction instruments owing to, among other things, their reduced sensitivity to the vacuum environment [37–40]. Were such an on-axis metal cathode to be usable as a conventional thermionic source in UEM, while also having the stability and performance seen for off-axis Ni Wehnelt-aperture photoemission, the need for different basic electron-beam alignments would be circumvented. Unfortunately, while the photoemission stability and usability is approximately comparable to that of the Ni Wehnelt aperture, stability as a thermionic source was quite poor. Accordingly, while it is informative to discuss the *pc*-Ta cathode performance as a photoemitter, our view is that the Ni Wehnelt aperture configuration is overall more appealing when considering a combination of convenience, stability, and usability. Indeed, additional complication for such a configuration consists only of needing two files instead of one for saved basic beam alignments.

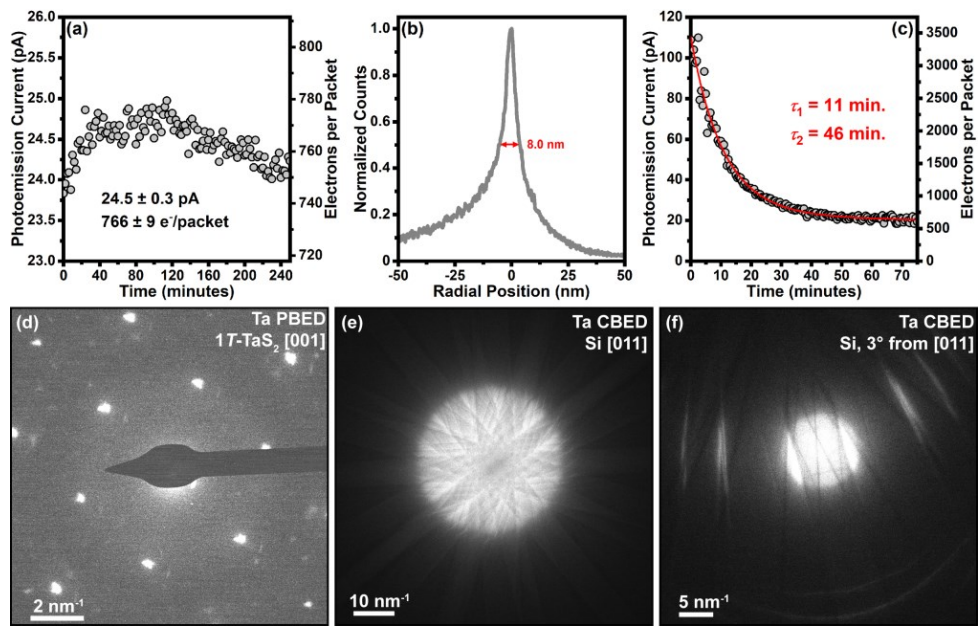

**Figure 6.** Photoemission stability, probe size, and diffraction-pattern quality from an on-axis, custom 200-μm diameter flat *pc*-Ta cathode. (a) Stability over a span of four hours of thermally equilibrated photoemission from the on-axis *pc*-Ta cathode. Here, a sub-thermionic-emission threshold heat-to value of 24 was used. Data acquisition was started 30 minutes after reaching this heat-to value. The average beam current and electrons per packet were 24.5 ± 0.3 pA and 766 ± 9 e⁻·packet⁻¹, respectively, for the UV laser settings used ($f_{rep}$ = 200 kHz). The error is one standard deviation from the mean. (Note the deviation from a steady beam current is an artifact, which could have been caused by a number of factors, such as a systematic variation in lab temperature and thus a systematic drift of optical harmonic conversion efficiencies.) (b) Optimized probe size generated from the *pc*-Ta cathode using the Nanoprobe mode of the Tecnai Femto. The probe size was 8 nm fwhm. Note that the asymmetric wings of the overall peak response arise from misalignment of the condenser aperture. (c) Photoemission stability of the *pc*-Ta cathode immediately following reduction of the heat-to value from 30 to 0. The data are fit with a bi-exponential decay function so that the decay times can be compared to the other sources and configurations tested. (d) PBED pattern of multilayer $1T$-TaS$_2$ along the [001] zone axis generated with the *pc*-Ta photoelectron beam. CBED patterns of Si (e) along the [011] zone axis and (f) approximately 3° from the [011] zone axis generated with the *pc*-Ta photoelectron beam and a uniform probe size of 28 nm fwhm.

Panels (a) and (b) in Figure 6 show the general stability and the optimized probe size, respectively, for photoemission from the on-axis *pc*-Ta cathode. Here, beam current was tracked for four hours continuously. For the specific data set shown in Figure 6a, one standard deviation from the mean was again ±1 %, similar to that for Wehnelt-aperture photoemission (see Figure 4a). (Note that the slight non-linear response is an artifact that likely arose from incomplete system equilibration.) The optimized probe size shown in Figure 6b is roughly half that of Wehnelt-aperture photoemission. However, this is because only a portion of the 200-μm diameter *pc*-Ta cathode was photoemitting [41]. Thus, the source size was substantially smaller than that of the Wehnelt aperture and the LaB$_6$ cathode. Indeed, probe sizes of less than 1 nm have been demonstrated in Schottky-FEG-based UEMs using side illumination of tungsten-needle sources with apex diameters of tens to ~100 nm [5]. Lastly, as with the other source materials and configurations, a bi-exponential fitting function was used to extract decay constants for photoelectron beam current generated from an initially hot and actively cooling *pc*-Ta cathode (Fig. 6c).

Finally, Figure 6d-f illustrates the basic usability of the *pc*-Ta photocathode. As with Wehnelt-aperture photoemitted beams, PBED patterns from 1*T*-TaS$_2$ (Fig. 6d) and CBED patterns from Si (Fig. 6e,f) were obtained. While the CBED patterns were of reasonable quality, the PBED patterns in particular suffered from the irregular shape of the *pc*-Ta photoemitting region. That is, while the relatively weak satellite peaks arising from the periodic lattice distortions associated with charge-density waves are observable, one can see the impact of the irregularly shaped emitting region in the Bragg-spot profiles. We consider this, however, to be only a minor issue, as one can readily translate the UV fs laser pulse train to other regions of the cathode in order to improve the spot shape [41]. Also, such irregularly shaped diffracted-beam profiles do not preclude application of ultrafast electron diffraction measurements. Rather, it is the poor thermionic performance of the *pc*-Ta cathode that makes it less desirable than Wehnelt-aperture photoemission for achieving one of the stated goals: stability and usability in both TEM and UEM modes.

**Summary and Conclusions**

We have focused on testing and comparing the stability and usability of three different cathode materials and electron-gun configurations for combined TEM and UEM operation: (1) on-axis thermionic and photoemission from a custom 100-µm diameter truncated LaB$_6$ with graphite guard ring, (2) off-axis photoemission from a 1-mm diameter Ni Wehnelt-aperture anode-facing surface, and (3) on-axis thermionic and photoemission from a custom 200-µm diameter truncated *pc*-Ta cathode. Overall, we found the combined stability and usability of the off-axis Ni Wehnelt-aperture photoemission for UEM to be superior to that of the other materials and configurations. Further, the off-axis configuration proves quite convenient for switching between UEM and TEM modes – an ideal LaB$_6$ source can be installed and used for TEM operation, while off-axis photoemission from the Ni Wehnelt-aperture surface avoids many of the challenges associated with using the same source for thermionic and photoemission. Future work on this configuration will include characterizing the shot-to-shot stability, measuring the photoelectron energy distribution [36], probing the photoemission mechanism (preliminary results indicate a single-photon photoemission mechanism) [42,43], and assessing the usability as a high spatiotemporal-resolution UEM source for real-space imaging of angstrom-femtosecond materials dynamics.

**Acknowledgments**

This material is based upon work supported by the U.S. Department of Energy, Office of Science, Office of Basic Energy Sciences under Award No. DE-SC0018204. This work was supported partially by the National Science Foundation through the University of Minnesota MRSEC under Award Number DMR-2011401. Part of this work was carried out in the College of Science and Engineering Characterization Facility, University of Minnesota, which has received capital equipment funding from the NSF through the UMN MRSEC program under Award Number DMR-2011401.

**Author Contributions**

**Simon Willis:** Data Curation; Formal Analysis; Investigation; Methodology; Validation; Visualization; Writing – original draft. **Wyatt Curtis:** Investigation; Methodology. **David Flannigan:** Conceptualization; Funding acquisition; Methodology; Project administration; Resources; Supervision; Visualization; Writing – review & editing.

**Author Declarations**

The authors have no competing interests to declare.